\documentclass[a4paper,fleqn,usenatbib]{mnras}

\usepackage{newtxtext,newtxmath}

\usepackage[T1]{fontenc}
\usepackage{ae,aecompl}

\usepackage{graphicx}	
\usepackage{amsmath}	
\usepackage{amssymb}	

\title[CH$_3$CN and CH$_3$CCH in IRAS 16293--2422]{Methyl cyanide (CH$_3$CN) and propyne (CH$_3$CCH) in the low mass protostar IRAS 16293--2422}

\author[Andron et al.]{
Ines Andron$^{1}$,
Pierre Gratier$^{1}$,
Liton Majumdar$^{2}$,
Thomas H. G. Vidal$^{1}$,
Audrey Coutens$^{1}$
\newauthor
Jean-Christophe Loison$^{3,4}$,
and Valentine Wakelam$^{1}$\thanks{E-mail: valentine.wakelam@u-bordeaux.fr}
\\
$^{1}$Laboratoire d'astrophysique de Bordeaux, Univ. Bordeaux, CNRS, B18N, all\'ee Geoffroy Saint-Hilaire, 33615 Pessac, France\\
$^{2}$Jet Propulsion Laboratory, California Institute of Technology, 4800 Oak Grove Drive, Pasadena, CA 91109, USA\\
$^{3}$Univ. Bordeaux, ISM, UMR 5255, F-33400, Talence, France\\
$^{4}$CNRS, ISM, UMR 5255, F-33400, Talence, France
}

\date{Accepted XXX. Received YYY; in original form ZZZ}

\pubyear{2018}

\begin{document}
\maketitle
%
\begin{abstract}
Methyl cyanide (CH$_3$CN) and propyne (CH$_3$CCH) are two molecules commonly used as gas thermometers for interstellar gas. They are detected in several astrophysical environments and in particular towards protostars. Using data of the low-mass protostar IRAS 16293--2422 obtained with the IRAM 30m single-dish telescope, we constrained the origin of these two molecules in the envelope of the source. The line shape comparison and the results of a radiative transfer analysis both indicate that the emission of CH$_3$CN arises from a warmer and inner region of the envelope than the CH$_3$CCH emission. 
We compare the observational results with the predictions of a gas-grain chemical model. Our model predicts a peak abundance of CH$_3$CCH in the gas-phase in the outer part of the envelope, at around 2000 au from the central star, which is relatively close to the emission size derived from the observations. The predicted CH$_3$CN abundance only rises at the radius where the grain mantle ices evaporate, with an abundance similar to the one derived from the observations. 

\end{abstract}




\section{Introduction}


Star-forming regions are ideal places for the development of the chemical complexity in the interstellar medium. The molecules detected in these regions can be used to better understand the network of interactions between all  present species that can lead to the formation of more complex molecules. All the species with 6 atoms or more and at least one atom of carbon, that are detected in the interstellar medium, are complex organic molecules (COMs, \citealt{Herbst2009}). The formation of these species  and  their origin, either from gas-phase or dust surface reactions, is still highly debated.

In addition to their chemical interest, molecules allow astronomers to constrain the physics of the studied regions. For example, some molecules probe shocks \citep[SO,][]{2001A&A...370.1017V,2015A&A...581A..85P} when others can be used to constrain the density or the temperature \citep[for instance HC$_3$N,][]{2000ApJ...542..870D}. In particular, symmetric top molecules are good indicators of the temperature of the source. Propyne (also called methyl acetylene, CH$_3$CCH) and methyl cyanide (CH$_3$CN) belong to this category \citep{Askne1984,Kalenskii2000}. 
These two molecules have been detected in a lot of environments: massive young stellar objects \citep[e.g.,][]{Fayolle2015}, low-mass star-forming regions \citep[e.g.,][]{vanDishoeck1995}, photodissociation regions \citep[e.g.,][]{Gratier2013,Guzman2014}, circumstellar envelopes of evolved stars \citep[e.g.,][]{Agundez2008,Agundez2015}, and even other galaxies \citep[e.g.,][]{Mauersberger1991}. Both molecules have also been detected in dense and cold cores \citep[e.g.,][]{Vastel2014,2016ApJS..225...25G}, while CH$_3$CN was also found towards a protoplanetary disk \citep{Oberg2015}.


In this study, we focus on the study of these two complex molecules towards the solar-type protostar IRAS 16293--2422 (hereafter IRAS16293), using data obtained with the IRAM-30m telescope. This deeply embedded source, located in the $\rho$ Ophiuchi cloud, is a binary composed of source A (South-East) and source B (North-West), which are separated by about 5$\arcsec$, i.e. about 750 au at a distance of about 141~pc \citep{Ortiz2017,2018arXiv180203234D}. This object is at a very early stage of the star formation process (Class 0, \citealt{Andre1993}) and has been characterized by the presence of numerous complex organic molecules in the warm inner regions of the two components, where the icy grain mantles thermally desorb \citep{Cazaux2003,Bottinelli2004,Bisschop2008,Jorgensen2011,Jorgensen2012,Jorgensen2016,Kahane2013,Lykke2017,Ligterink2017}. A more complete description of this source is presented in \citet{Jorgensen2016}.

Using the 30m IRAM single dish antenna, we have carried out a 16 GHz spectral survey of the source. Based on these observations and using similar radiative transfer analysis, we have published, in the past, studies about CH$_3$SH \citep{2016MNRAS.458.1859M}, c-C$_3$HD \citep{2017MNRAS.467.3525M}, and HOCO$^+$ \citep{2018arXiv180305442M}. In this paper, we present a consistent analysis of CH$_3$CCH and CH$_3$CN. This paper is organized as follows. In Section 2, we present the data, their analysis, and the observational results. Section 3 includes a description of the chemical model and its results in comparison with the observations. Last, we conclude in Section 4.


\section{Observations and data reduction}

\subsection{Observations}

Observations were performed using the IRAM 30m telescope from August 18 to 23, 2015 in average summer conditions (a median value of 4-6mm water vapor). The EMIR heterodyne 3mm receiver tuned at a frequency of 89.98 GHz was used in the Lower Inner sideband and paired with the Fourier Transform Spectrometer in its 195 kHz resolution mode. The observed spectrum is composed of two approximately 8 GHz regions centered respectively on 88.41 and 104.06 GHz. The typical angular resolution is 24-28$''$. The wobbler switching mode with a throw of 90" and a period of 2 seconds was used to make observations centered at the position $\alpha$2000 = 16h32m22.75s, $\delta$2000 = -24$^o$28'34.2" , midway between sources A and B of IRAS16293. This throw ensures a flat baseline even for observations at low elevation and in summer conditions. Moreover, at the beginning of each run and after sunset, the nearby planet Saturn was used for focus. Pointing was checked hourly on nearby quasars with a pointing correction less than a third of the beam.

\subsection{Line properties of CH$_3$CN and CH$_3$CCH}\label{line_prop}

\begin{table*}
	\centering
	\caption{Spectroscopic parameters and observed line properties.}
	\label{table_obs}
	\begin{tabular}{llllllllllll} 
		\hline
		Molecules  & QNs                 	& Frequency  & A$_{i,j}$ 			 & E$_{up}$ &  g$_{up}$ &   T$_{\rm peak}$ & $\sigma$ & Area & $\varv_{\rm LSR}$ & FWHM \\
		           & $J_K \rightarrow J'_K$ & (MHz) 	 & (s$^{-1}$) 			 & (K)      &           & (mK) & (mK) & (K km s$^{-1}$) & (km s$^{-1}$) & (km s$^{-1}$)\\
		\hline                                                                                 
		CH$_3$CN  & $5_4\rightarrow 4_4$ 	& 91958.726  & $2.28 \times 10^{-5}$ & 127.5    &  22       & 36.2 & 1.93& 0.312$\pm$0.005 &  3.0$\pm$0.3 & 8.3$\pm$0.2 \\
		CH$_3$CN  & $5_3\rightarrow 4_3$ 	& 91971.130  & $4.05 \times 10^{-5}$ & 77.5     &  44       & 91.2 & 3.07& 0.68$\pm$0.01  & 2.95$\pm$0.05 & 7.0$\pm$0.1\\
        CH$_3$CN  & $5_2\rightarrow 4_2$ 	& 91979.994  & $5.32 \times 10^{-5}$ & 41.8     &  22       & 86.5 & 2.86 & 0.553$\pm$0.008 & 3.00$\pm$0.05& 6.0$\pm$0.1 \\
		CH$_3$CN  & $5_1\rightarrow 4_1$ 	& 91985.314  & $6.08 \times 10^{-5}$ & 20.4     &  22       & 113 & 2.13 & 0.679$\pm$0.008 & 3.40$\pm$0.03 & 5.60$\pm$0.07\\
		CH$_3$CN  & $5_0\rightarrow 4_0$ 	& 91987.088  & $6.33 \times 10^{-5}$ & 13.2     &  22       & 115 & 2.09 & 0.672$\pm$0.001 & 2.92$\pm$0.01 & 5.49$\pm$0.05\\
		\hline                           	                                                   
		CH$_3$CCH & $5_4\rightarrow 4_4$ 	& 85431.174  & $7.30 \times 10^{-7}$ & 127.9    &  22       & $< 7.32$ & 2.44 & - & - & - \\
		CH$_3$CCH & $5_3\rightarrow 4_3$ 	& 85442.601  & $1.30 \times 10^{-6}$ & 77.3     &  44       & 2.63 & 2.51 & 0.090$\pm$0.006 & 3.5$\pm$0.1 & 3.2$\pm$0.3\\
		CH$_3$CCH & $5_2\rightarrow 4_2$ 	& 85450.766  & $1.70 \times 10^{-6}$ & 41.2     &  22       & 70.9 &  2.11 & 0.145$\pm$0.004 & 3.62$\pm$0.02 & 1.92$\pm$0.06\\
		CH$_3$CCH & $5_1\rightarrow 4_1$ 	& 85455.667  & $1.95 \times 10^{-6}$ & 19.5     &  22       & 207 & 2.83 & 0.398$\pm$0.005 & 3.62$\pm$0.01 & 1.78$\pm$0.03 \\
		CH$_3$CCH & $5_0\rightarrow 4_0$ 	& 85457.300  & $2.03 \times 10^{-6}$ & 12.3     &  22       & 267 & 2.72 & 0.497$\pm$0.004 & 3.691$\pm$0.008 & 1.75$\pm$0.02 \\
		CH$_3$CCH & $6_5\rightarrow 5_5$ 	& 102499.019 & $1.09 \times 10^{-6}$ & 197.8    &  26       & $< 7.38$& 2.46  & - & - & - \\
		CH$_3$CCH & $6_4\rightarrow 5_4$ 	& 102516.637 & $1.98 \times 10^{-6}$ & 132.8    &  26       & $< 6.48$ & 2.16 & - & - & - \\
		CH$_3$CCH & $6_3\rightarrow 5_3$ 	& 102530.348 & $2.67 \times 10^{-6}$ & 82.3     &  52       & 43.2 &  1.29  & 0.13$\pm$0.03 & 3.7$\pm$0.3 & 2.8$\pm$0.7 \\
		CH$_3$CCH & $6_2\rightarrow 5_2$ 	& 102540.145 & $3.16 \times 10^{-6}$ & 46.1     &  26       & 98.5 &  7.16 & 0.2$\pm$0.1 & 3.8$\pm$0.6 & 2.4$\pm$0.8 \\
		CH$_3$CCH & $6_1\rightarrow 5_1$ 	& 102546.024 & $3.46 \times 10^{-6}$ & 24.5     &  26       & 218 & 5.85 & 0.6$\pm$0.1 & 3.8$\pm$0.1 & 2.1$\pm$0.4 \\
		CH$_3$CCH & $6_0\rightarrow 5_0$ 	& 102547.984 & $3.56 \times 10^{-6}$ & 17.2     &  26       & 262 & 4.96 & 0.71$\pm$0.08 & 3.8$\pm$0.1 & 2.1$\pm$0.3 \\
		\hline
	\end{tabular}
	\\
	Notes: QNs: quantum numbers, Frequency: transition rest frequency, A$_{i,j}$: Einstein coefficient of spontaneous emission of a photon by transition from level $j$ to level $i$, E$_{up}$: energy of the upper level, g$_{up}$: statistical degeneracy of the upper level, T$_{\rm peak}$: peak observed main beam temperature, $\sigma$: observed noise, Area: observed integrated intensity, $\varv_{\rm LSR}$: observed doppler velocity shift, FWHM: observed line full width at half maximum.
\end{table*}

\begin{table*}
	\centering
	\caption{Prior distribution functions for the parameters used in the Bayesian approach.}
	\label{prior_dist}
	\begin{tabular}{llll} 
		\hline
		Name & Parameter & Distribution & Comment \\
		\hline
		Molecular column density & $\log$~N (cm$^{-2}$) & Uniform(8,22) \\
		Excitation temperature & T$_{\rm ex}$ (K) & Normal(40,40) & limited to T>2.73K\\
		Radius of the emission & $\log$ R (au) & Normal($\log(R(T))$,0.01) & $R(T)$ is the relationship from Crimier et al. 2010 \\
		Doppler shift & $\varv_{LSR}$ (km s$^{-1}$) & Uniform(2.8,4.8)\\
	        FWHM of the emission & $\Delta \varv$ (km s$^{-1}$) & Normal(5,5)\\
		Additional noise & $\sigma_{\rm add}$ (mK) & Normal(0,1)\\
		\hline
	\end{tabular}
	\\
	Notes: Uniform($x_l$,$x_u$) is the uniform random distribution that can take values between $x_l$ and $x_u$, Normal($\mu$,$\sigma$) is the normal (gaussian) random distribution with mean $\mu$ and standard deviation $\sigma$.
	
\end{table*}

\begin{table}
	\centering
	\caption{Point estimates of the posterior distribution function corresponding to the median and one sigma uncertainty for CH$_3$CN and CH$_3$CCH.}
	\label{posterior_dist}
	\begin{tabular}{lll} 
		\hline
		Parameter & CH$_3$CCH & CH$_3$CN \\
		\hline
		$\log$~N (cm$^{-2}$) &14.73$\pm$0.03 & 16.14$\pm$ 0.05 \\
		T$_{\rm ex}$ (K) & 25$\pm$1 & 75$\pm$2 \\
        $\varv_{LSR}$ (km s$^{-1}$) &  3.79$\pm$0.01 & 3.18$\pm$0.06 \\
		$\theta ('')$ & 11.7$\pm$0.5 & 1.19$\pm$0.04 \\
		$\Delta \varv$ (km s$^{-1}$) & 1.99$\pm$0.02 &  5.1$\pm$0.1 \\
		$\log [X]^a$ & -9.63$\pm$0.03 &  -8.10$\pm$0.06 \\
		$\log N_{H_{tot}}$$^b$ & 24.36$\pm$0.002 & 24.26$\pm$0.03\\
		\hline
	\end{tabular}
    \linebreak
    Notes: $\theta ('')$ is the source size, $^a [X] = N_x/N_{H_{tot}}$, $^b$ $N_{H_{tot}}$ is the total hydrogen nucleon density.
\end{table}

In Table~\ref{table_obs}, we provide some line properties. The spectroscopic data, extracted from the CDMS database \citep{2005JMoSt.742..215M}, are from \citet{Muller.2015} for CH$_3$CN and \citet{Cazzoli.2008} for CH$_3$CCH.
We used the CLASS software, from the GILDAS package\footnote{http://www.iram.fr/IRAMFR/GILDAS}, to reduce and analyse the data. Gaussian fits were made to the detected lines following a local low (typically 0th) order polynomial baseline subtraction. Table~\ref{table_obs} shows the results of these fits for the 5 observed lines of CH$_3$CN and 8 observed lines of CH$_3$CCH. Three CH$_3$CCH observed lines are below the detection level and we report the 3$\sigma$ detection limit in Table~\ref{table_obs}. For CH$_3$CCH, the mean LSR velocity is 3.7 km s$^{-1}$ and the mean FWHM is 2.3 km s$^{-1}$ while for CH$_3$CN, the LSR velocity is 3.1 km s$^{-1}$ and the mean FWHM is 6.5 km s$^{-1}$. 

\subsection{CH$_3$CN and CH$_3$CCH radiative transfer modeling}

We use a Bayesian approach similar to the one presented in \citet{2016MNRAS.458.1859M} to recover the distribution of parameters. The radiative transfer modeling is carried out assuming Local Thermal Equilibrium. In addition to the fitting uncertainties, we allowed for a 10\% calibration error. The modelled line profiles are computed as a function of the molecule column density, excitation temperature, line width, systemic velocity, and source size (described by a two dimensional circular Gaussian source profile centered in the telescope beam). The line shapes are computed assuming a Gaussian opacity profile as a function of frequency. In our computation, we have used the relation between the source size and the temperature profile determined by \citet{2010A&A...519A..65C} (see their Fig. 7). \citet{2010A&A...519A..65C} have determined the density and temperature profiles (at large spatial scale) of the IRAS16293 protostellar envelope using continuum data (single dish and interferometric, from millimeter to MIR) and ISO water observations. They showed that the densities remain higher than $10^6$ cm$^{-3}$ even at a distance of 2000 au from the two central objects. This is higher than the typical critical densities for the observed transitions of CH$_3$CN that can be computed using the available collisional coefficients from \citet{1986ApJ...309..331G}, that cover a range between $8\times10^4$ cm$^{-3}$ and $5\times10^5$ cm$^{-3}$ at 60K. The LTE approximation should remain valid for our study. Collisional coefficients for CH$_3$CCH are not known.

The sampling of the posterior distribution function was carried out using the No-U-Turns (NUTS) Hamiltonian Monte Carlo sampler implemented in the Stan Probabilistic Programming Language \citep{JSSv076i01} with the Pystan\footnote{Stan Development Team. 2017. Pystan: the Python interface to Stan, Version 2.17.1 \url{http://mc-stan.org}} interface. Four independent chains were run for 4000 iterations, discarding the first half for burn-in and the adaptation of the Hamiltonian Monte Carlo NUTS parameters. Convergence was checked by computing the Gelman-Rubin $\hat{R}$ test \citep{Gelman.1992} ensuring that the values were below 1.01 for all parameters. The properties of the prior distributions are given in Table~\ref{prior_dist}.

\begin{figure*}
\includegraphics[width=1\linewidth]{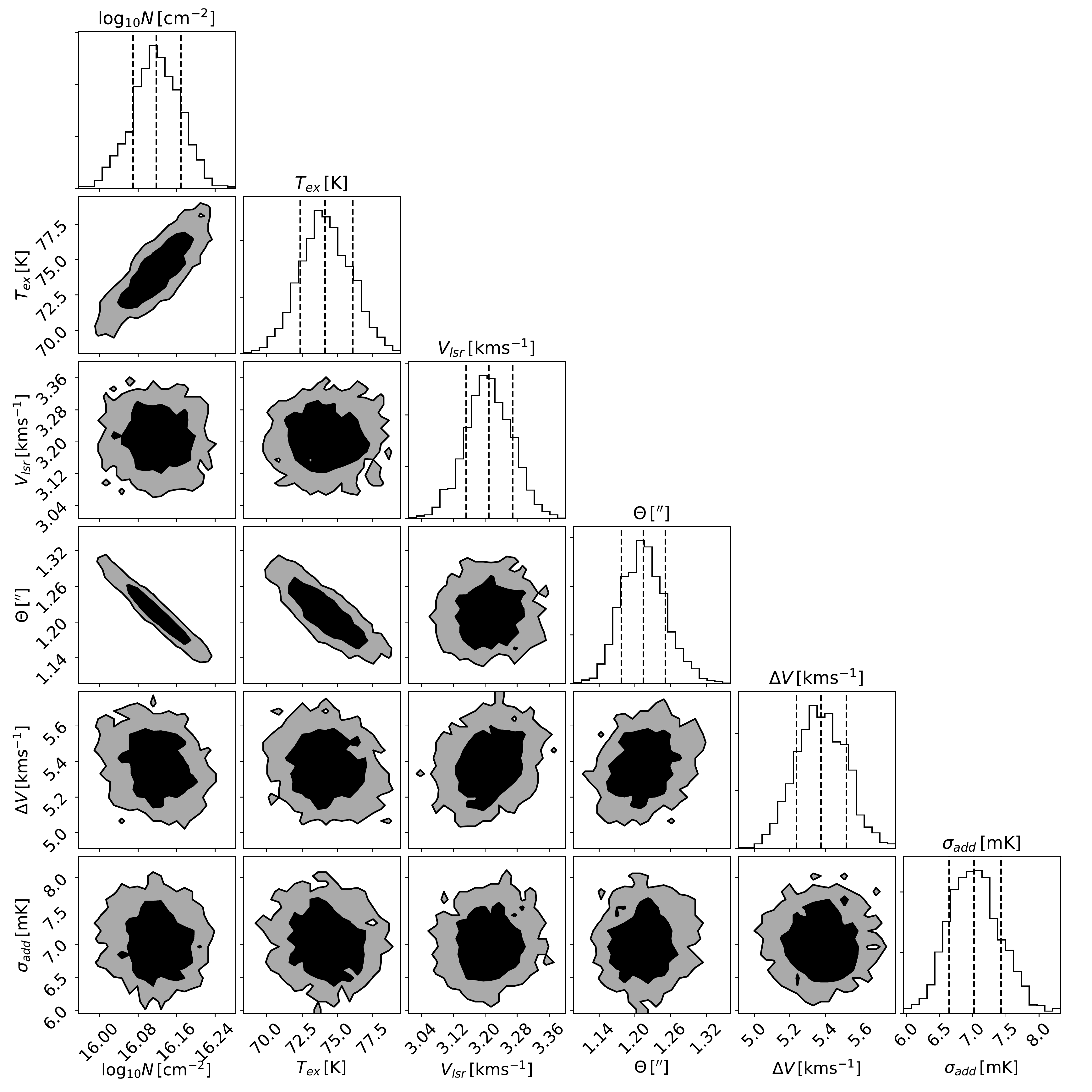}
\caption{One-dimensional and two-dimensional (i.e. contour) histograms of the posterior distribution of parameters for CH$_3$CN. Contours contain 68\% and 95\% of samples, respectively. Vertical dashed lines correspond to 16\%, 50\%, and 84\% of the samples.
\label{CH3CN_corner}}
\end{figure*}

\begin{figure*}
\includegraphics[width=1\linewidth]{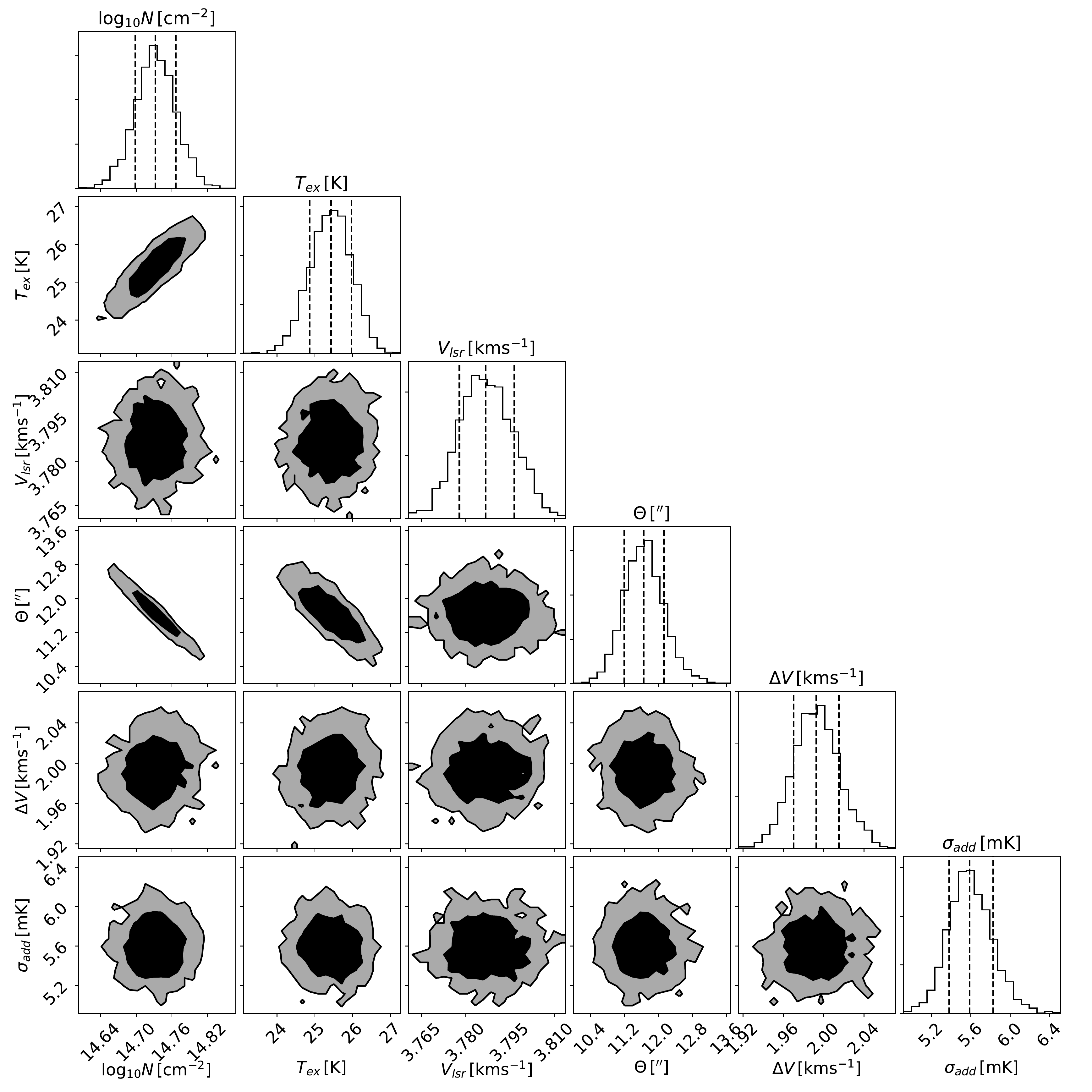}
\caption{Same as Fig.~\ref{CH3CN_corner} but for CH$_3$CCH.
\label{CH3CCH_corner}}
\end{figure*}

\begin{figure*}
\includegraphics[width=1\linewidth]{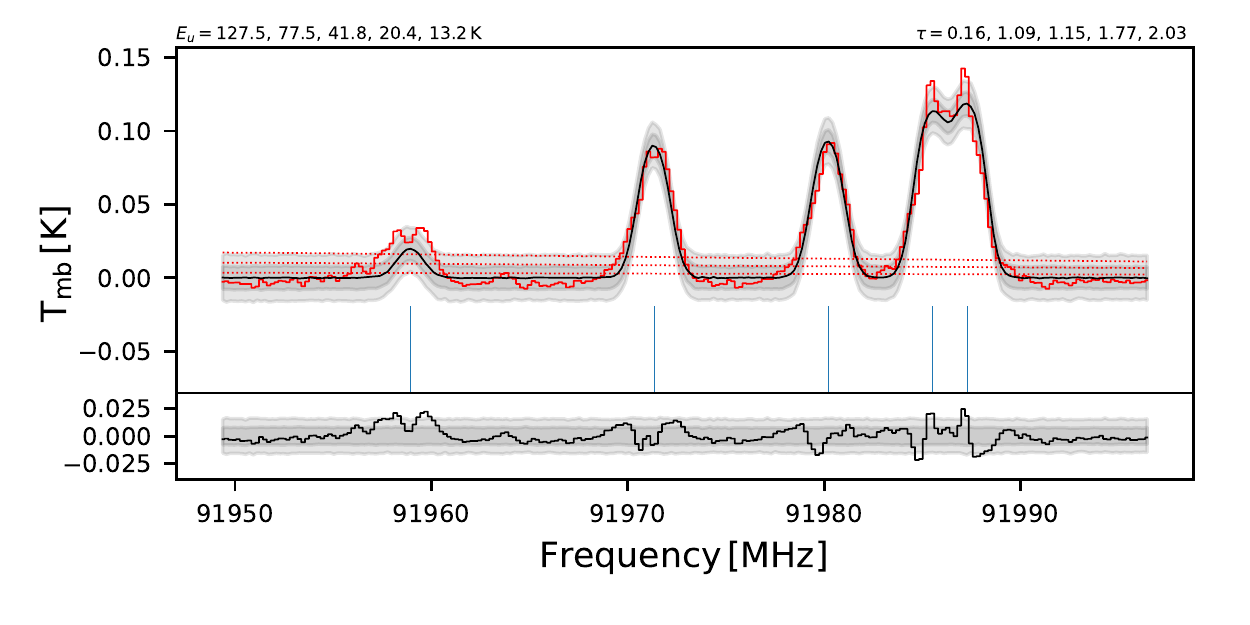}
\caption{Upper panel: observed (red) and modelled spectra for CH$_3$CN. The modelled spectra are represented by the median of the distribution of models (solid line) along with the associated $1\sigma$ (dark grey) and $2\sigma$ (light grey) uncertainties, the thin  vertical blue lines are the frequencies of the fitted lines,  the red dotted lines correspond to the $1\sigma$, $3\sigma$, and $5\sigma$ noise level. Lower panel: distribution of the residuals plotted as the median (solid line) along with the associated $1\sigma$ (dark grey) and $2\sigma$ (light grey) uncertainties. Top left: values of the upper energy levels of the lines from left to right, top right:  median values of the inferred opacities for the lines.
\label{CH3CN_spectrum}}
\end{figure*}

\begin{figure*}
\includegraphics[width=1\linewidth]{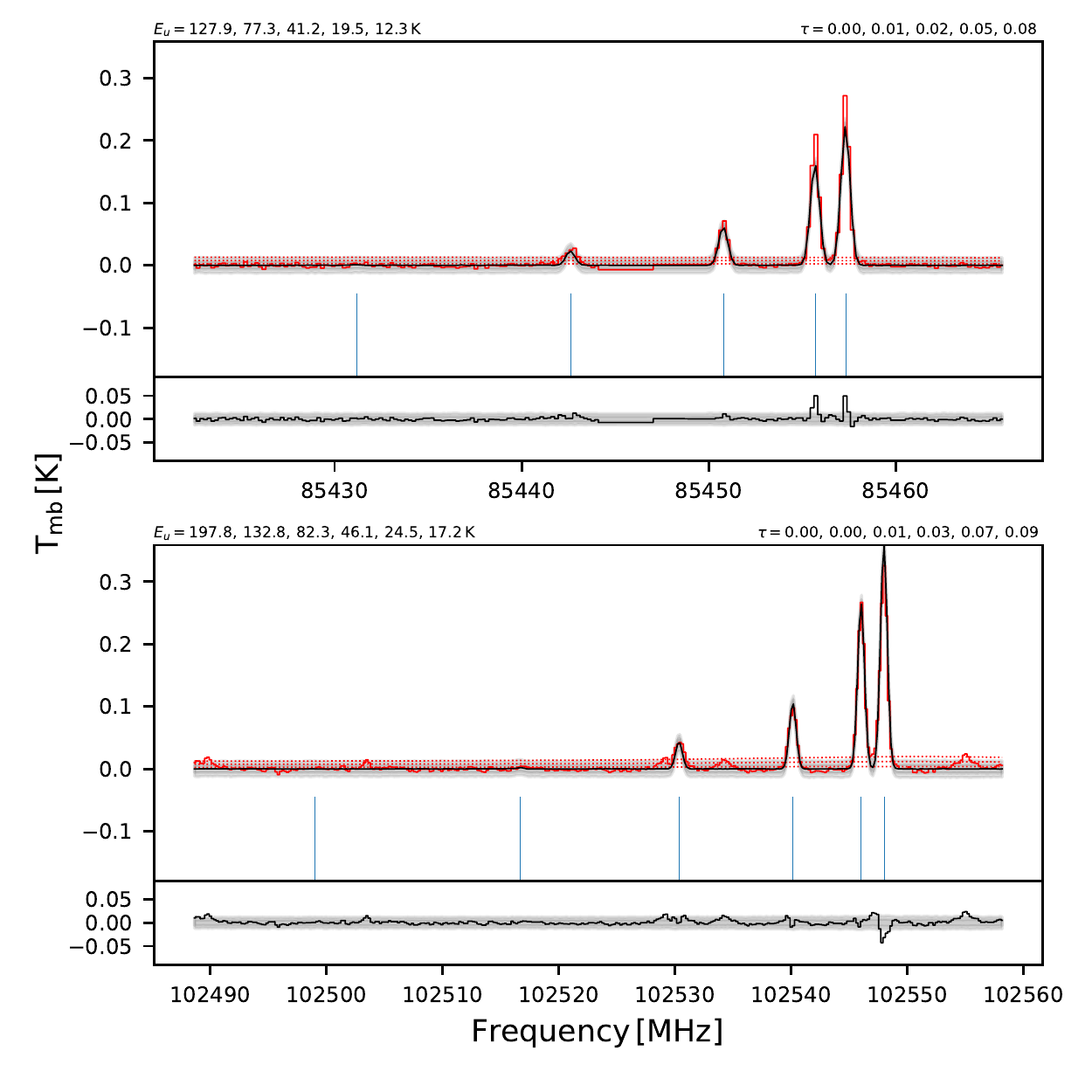}
\caption{Same as Fig.~\ref{CH3CN_spectrum} but for CH$_3$CCH.
\label{CH3CCH_spectrum}}
\end{figure*}

Figures \ref{CH3CN_corner} and \ref{CH3CCH_corner} show the 1D and 2D histograms of the posterior probability distribution function and the comparison of the observations with the distribution of computed intensities corresponding to the posterior distribution of parameters.
The summary of the point estimates for both molecules is presented in Table~\ref{posterior_dist}. Integrating the density power law from \citet{2010A&A...519A..65C} up to the source size emission (itself determined by comparing the excitation temperature of the molecules to the temperature profile from Crimier et al.) derived by the Bayesian method allows us to get an estimate of the H$_2$ column density of the emitting zone. The derived H$_2$ column densities are $2.3\times 10^{24}$ cm$^{-2}$ for CH$_3$CCH and $1.8\times 10^{24}$ cm$^{-2}$ for CH$_3$CN.
 The observed and modelled spectra for the four lines of CH$_3$CN and the 8 detected lines of CH$_3$CCH are displayed in Figs.~\ref{CH3CN_spectrum} and \ref{CH3CCH_spectrum}. The features not fitted by the model in Fig.~\ref{CH3CN_spectrum} probably arise from second order effects of radiative transfer (i.e. self absorption)  through the colder envelope that cannot be modelled by our 0D approach.

\subsection{Results}

According to the classification proposed by \citet{2011A&A...532A..23C}, both species are of kinematic type IV, meaning that their emission probably comes from both components of the source (A and B), and the common envelope. The rest velocities and the line widths of the CH$_3$CCH lines are similar to those of molecules probing the cold envelope of the protostellar system \citep[v$_{\rm LSR}$ $\sim$3.9 km s$^{-1}$ and FWHM$\sim$2 km s$^{-1}$,][]{2011A&A...532A..23C}. The CH$_3$CN lines, on the contrary, present larger widths $>$ 5.5 km s$^{-1}$.  The computed excitation temperatures are different for these two molecules and much higher for CH$_3$CN (75~K) as compared to CH$_3$CCH (25~K). These results seem to indicate that the CH$_3$CN emission comes from a warmer region, probably associated with the hot corino(s), while the CH$_3$CCH emission comes from the colder outer envelope.  Based on the excitation temperatures and the radial temperature profile used, the CH$_3$CN emission would come from about 170 au from the central star whereas CH$_3$CCH would come from about 1700 au (assuming the most recent distance of the source, i.e. 141 pc). Integrating the total hydrogen nucleon density column density from \citet{2010A&A...519A..65C} within these radii, we obtain abundances of $2\times 10^{-10}$ for CH$_3$CCH and $8\times 10^{-9}$ for CH$_3$CN (with respect to the total hydrogen nucleon density).  In addition, the structure determined by \citet{2010A&A...519A..65C} has to be taken with caution as the authors assumed a distance of 120 pc for the source while this value has recently been revisited to a larger one  \citep[i.e. 141~pc,][]{Ortiz2017,2018arXiv180203234D}.\\
 The structure of IRAS16293 used in the paper is assumed to be spherical, although it is a binary source with a complex structure in the inner regions \citep{Jorgensen2016,2018A&A...612A..72J}. The dust properties are also very uncertain in these regions. This has consequences on the derived abundance with respect to H$_2$. If the molecular emission comes from large scales, the abundance with respect to H$_2$ should be relatively well constrained while the abundances in the inner regions (within a few arcsec) would be more uncertain. The absolute abundance of CH$_3$CCH should consequently be better constrained than the one of CH$_3$CN. If the emission of the molecule is only at a certain radius of the envelope, the derived abundances should be considered as upper limits.
Recent ALMA analysis of CH$_3$CN show that this molecule in indeed emitting towards both protostellar sources and does not show any significant difference between the two sources. \citet{2018A&A...616A..90C}  obtain a very similar column density (within a factor 2) for the two components. If we correct the column densities determined in \citet{2018A&A...616A..90C} by the source size constrained in this paper, the column density we obtain is in fact the average of the column densities of the two sources, which confirms that our analysis is correct. For the CH$_3$CCH molecule, since the emission is much more extended, the values derived here should be less biased by the multiplicity of the source.

\section{Comparison with chemical models}

\subsection{Model description}

To simulate the chemistry in the envelope of IRAS16293, the 3-phase Nautilus gas-grain code has been used \citep{2016MNRAS.459.3756R}. This numerical model computes the gas and ice composition as a function of time by solving a set of differential equations, which relate the species abundances to the chemical rates. 
In addition to the gas-phase reactions \citep[see][]{2015ApJS..217...20W}, interactions between the gas-phase species and the grain surfaces are included: physisorption of gas-phase species onto the grains and thermal and non thermal desorption. For the non thermal desorption, cosmic-ray induced desorption \citep{1993MNRAS.261...83H}, photodesorption \citep[see][]{2016MNRAS.459.3756R}, and chemical desorption \citep{2007A&A...467.1103G} are included. Reactions at the surface of the grains follow the Langmuir-Hinshelwood theory. All the parameters for the surface chemistry are the same as in \citet{2016MNRAS.459.3756R} while the gas and surface chemical networks are the same as in \citet{2017MNRAS.469..435V}. 

Using this code, the chemistry is then computed in cells of material falling into the centre of the protostar. The physical parameters (temperatures, densities, and visual extinctions) are the results of radiation hydrodynamical simulations from \citet{2000ApJ...531..350M}.
This structure has already been used in several previous studies of this source \citep{2008ApJ...674..984A,2012ApJ...760...40A,2014MNRAS.445.2854W,2014MNRAS.441.1964B,2016MNRAS.458.1859M,2017MNRAS.467.3525M}. The time dependent density and temperature profiles are shown in Fig. 2 of \citet{2008ApJ...674..984A}.
 As already discussed in \citet{2014MNRAS.445.2854W}, the physical structure at the end of the hydrodynamical simulations is similar to the temperature and density gradients derived by \citet{2010A&A...519A..65C} from the observations for IRAS16293 \citep[see Fig. 1 of][ for a comparison]{2014MNRAS.445.2854W}. The density structure is however approximately ten times smaller than the observed one. As in previous studies, we then multiply all the densities of the physical model (at all times and all radii) by a factor of ten to be closer to the observations. The physical model is then not self-consistent anymore. However, in the absence of a time-dependent physical model reproducing the exact observed structure, we decided to use this one because the effect of the dynamics has a major impact on the chemical structure of the envelope \citep{2018MNRAS.474.5575V}. See \citet{2014MNRAS.445.2854W} for a complete discussion on this point.
 
We use as initial abundances for this dynamical model the output of a dark cloud chemical simulation. The physical parameters of this  initial simulation are: a gas and dust temperature of 10 K, a total hydrogen nucleon density of $2\times10^4$ cm$^{-3}$, a cosmic-ray ionization rate of $1.3\times10^{-17}$ s$^{-1}$, and a visual extinction of 15 mag. The set of elemental abundances we used is summarized in Table \ref{tab_4}. We start with all species in their atomic or ionized form, with the exception of hydrogen, which is assumed to be entirely in its molecular form. \citet{2017MNRAS.469..435V} showed that the Nautilus chemical model does not require additional depletion of sulphur from its cosmic value in order to reproduce dark clouds observations, we therefore use it as the initial sulphur abundance. The final chemical composition obtained for a cloud age of 10$^6$ yrs is then used as initial conditions for the collapsing source. The choice of the initial cloud age is always a difficult one as the chemical modeling result may depend on this. With the dynamical physical structure used here, \citet{2018MNRAS.474.5575V} however showed that the model predictions do not depend much on the cloud age. We tested with a younger cloud of $10^5$~yrs and this is indeed the case for CH$_3$CCH and CH$_3$CN.

 After running the chemical model for the different infalling cells of material, we reconstruct the final chemical composition of the protostellar envelope in 1D, and this is what is shown in the next section.

\begin{table}
\caption{Elemental abundances used for the dark cloud run. *$a(b)$ stands for $a\times10^b$}
	\begin{center}
		\begin{tabular}{l r r}
		\hline
		\hline
   		Element & $n_i/n_H$* & References \\
   		\hline
		H$_2$    & 0.5             &    \\
   		He          & 0.09           & 1 \\
		N            & 6.2(-5)       &  2 \\
		O            & 2.4(-4)       &  3 \\
		C$^+$    & 1.7(-4)       &  2 \\
		S$^+$    & 1.5(-5)       &  2 \\
		Si$^+$   & 8.0(-9)       &  4 \\
		Fe$^+$  & 3.0(-9)       & 4\\
		Na$^+$  & 2.0(-9)       & 4 \\
		Mg$^+$ & 7.0(-9)       & 4\\
		P$^+$   & 2.0(-10)     & 4\\
		Cl$^+$  & 1.0(-9)       & 4 \\
		F           & 6.7(-9)     & 5 \\
		\hline
 		\end{tabular}
	\end{center}
	\medskip{(1) \citet{2008ApJ...680..371W}, (2) \citet{2009ApJ...700.1299J}, (3) \citet{2011A&A...530A..61H}, (4) Low-metal abundances from \citet{1982ApJS...48..321G}, (5) Depleted value from \citet{2005ApJ...628..260N}}
  	\label{tab_4}
\end{table}

\subsection{Chemical model results}

\begin{figure}
\includegraphics[width=1\linewidth]{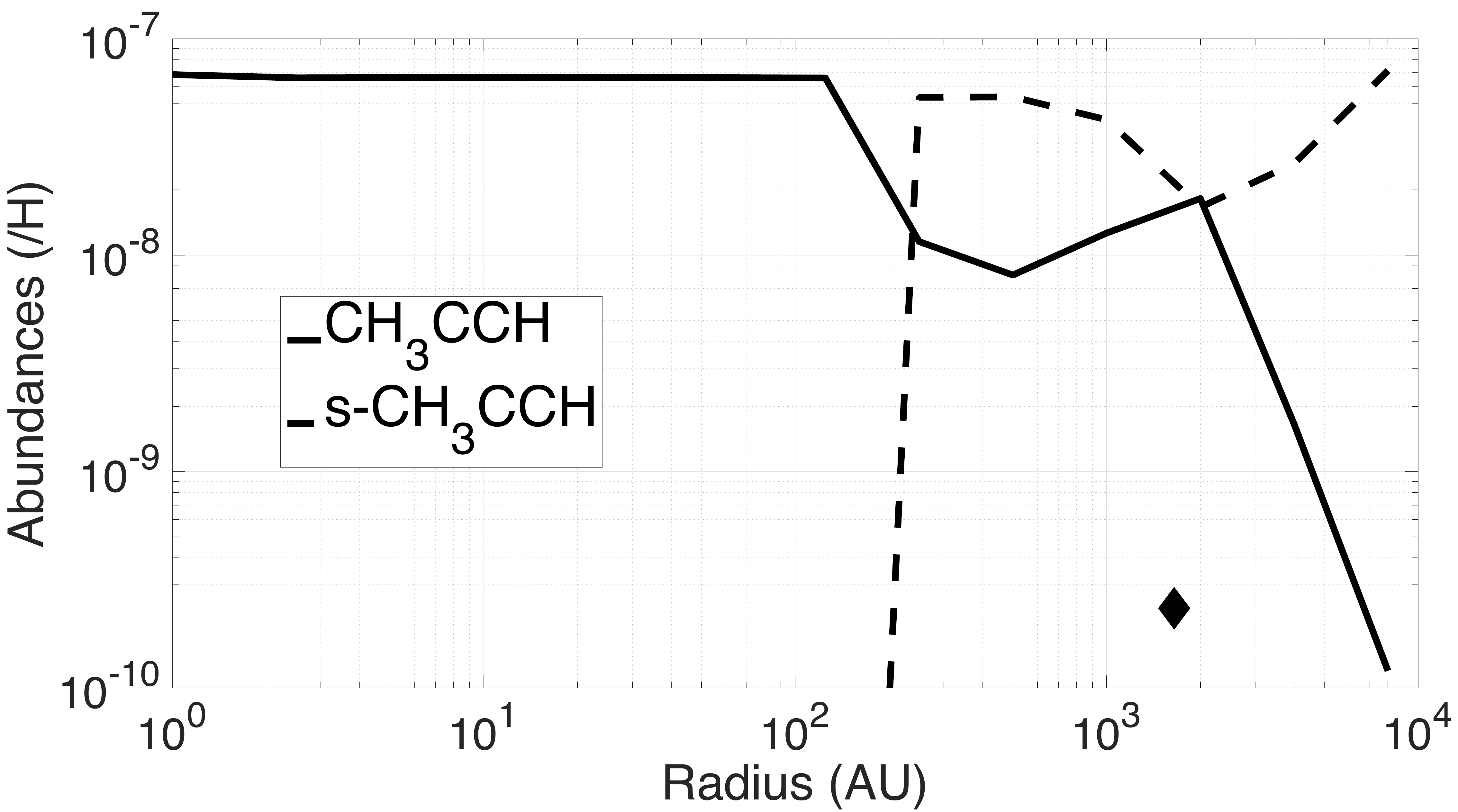}
\caption{Solid (dotted line) and gas-phase (solid line) CH$_3$CCH abundances predicted by the chemical model at the end of the simulations as a function of radius to the centre of the protostar. The diamond represents the observed abundance at the expected location. \label{CH3CCHab}}
\end{figure}

\begin{figure}
\includegraphics[width=1\linewidth]{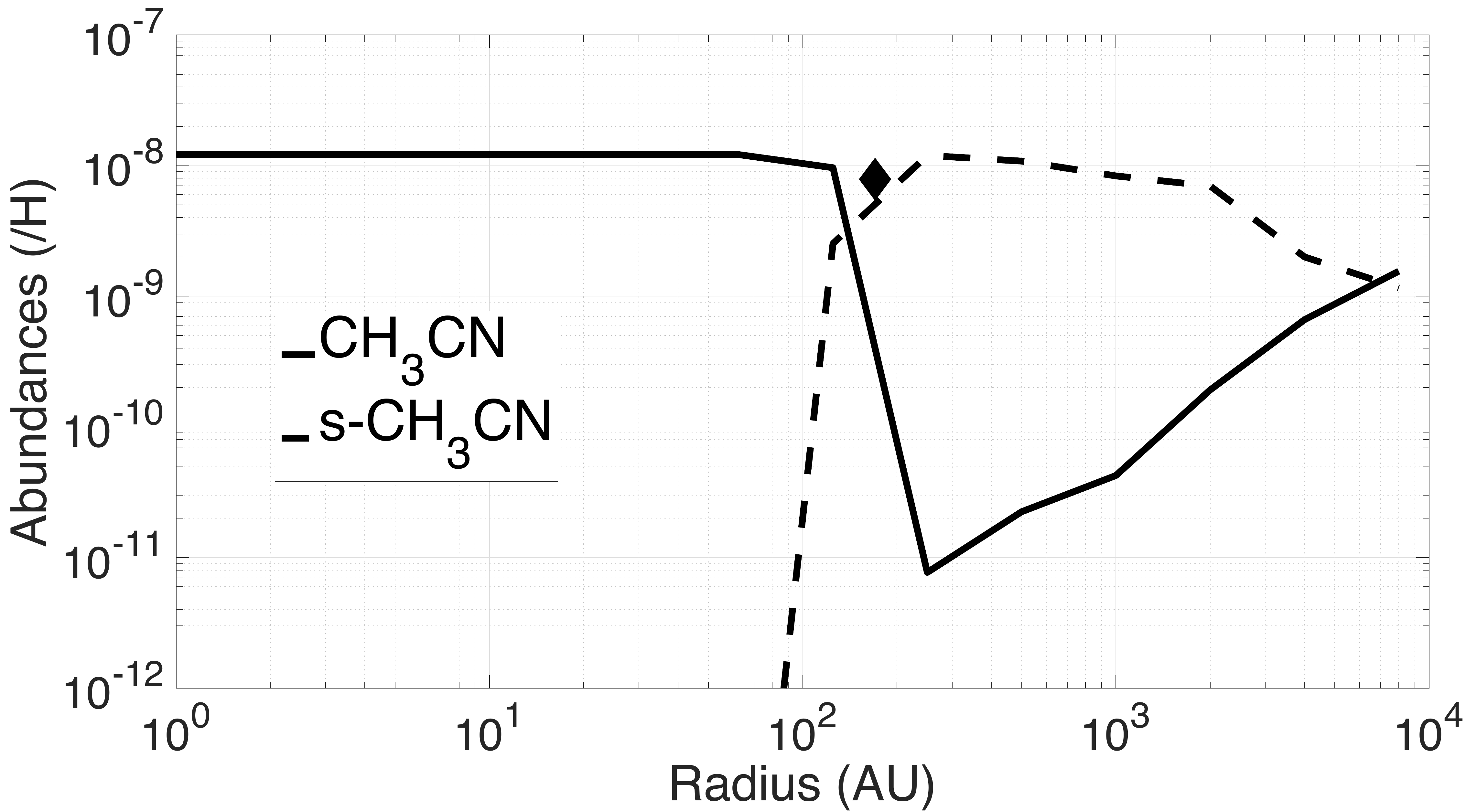}
\caption{Solid (dotted line) and gas-phase (solid line) CH$_3$CN abundances predicted by the chemical model at the end of the simulations as a function of radius to the centre of the protostar. The diamond represents the observed abundance at the expected location.  \label{CH3CNab}}
\end{figure}

Figures \ref{CH3CCHab} and \ref{CH3CNab} show solid and gas-phase abundances of CH$_3$CCH and CH$_3$CN respectively computed by the chemical model at the end of the protostellar simulations (i.e. at $3.43\times 10^5$~yr after the beginning of the collapse). For both molecules, the solid phase abundance at radii larger than 200 au is higher than the gas-phase one showing that both molecules are probably efficiently produced on the grains at low temperature. Indeed CH$_3$CCH is formed on the grain surface through successive hydrogenation of physisorbed C$_3$ \citep{2016MolAs...3....1H}, by:

\begin{eqnarray}
\text{C$_3$} \xrightarrow{\text{H}} \text{c-C$_3$H} \xrightarrow{\text{H}} \text{c-C$_3$H$_2$} \xrightarrow{\text{H}} \text{CH$_2$CCH} \xrightarrow{\text{H}} \text{CH$_3$CCH} \label{eq_2}
\end{eqnarray}

 In this model, the large abundance of C$_3$ is due to various efficient production pathways in the gas-phase associated with an absence of efficient destruction reactions as C$_3$ does not react quickly with H, N, C or O atoms \citep[see][]{2016MolAs...3....1H}.

CH$_3$CN is also efficiently formed on the grain surface through the hydrogenation of adsorbed H$_2$CCN, which is originally formed in the gas phase by:

\begin{eqnarray}
	\text{CN} + \text{CH$_3$} \to \text{H} + \text{H$_2$CCN} \label{eq_3}
\end{eqnarray}

For both species, their solid-state abundances drop sharply around 200 au because they are evaporated from the surfaces as the cells of material are moving inward. The binding energies used in the model are 3800 K for CH$_3$CCH and 4680 K for CH$_3$CN \citep{2017MolAs...6...22W} so that the evaporation temperature is slightly higher for CH$_3$CN and the molecule desorbs closer to the protostars. The evaporation radius depends obviously on the assumed binding energies. Recent experimental results on the CH$_3$CN binding energies on water ices by \citet{2017A&A...598A..18B} gave a mean binding energy of 6150~K, i.e. much larger than what we have used for this work. Using such value would put the evaporation radius of this molecule closer to the central star but would not change the abundance values as it would not change much its diffusion on the surface. This conclusion would also apply to CH$_3$CCH, i.e. a change of binding energy for this species would just change the evaporation radius. It should however be noted that the new binding energy of CH$_3$CN should be taken with great care as noted in \citet{2017A&A...598A..18B} because of the co-desorption of water with CH$_3$CN during the experiment. \\
The predicted gas-phase abundance of CH$_3$CCH in the outer envelope is quite high compared to the one of CH$_3$CN. Indeed, in the cold envelope (T < 20 K), CH$_3$CCH is also efficiently formed in the gas-phase from the hydrocarbons C$_2$H$_4$ and C$_3$H$_5$ produced in the parent cold cloud via the following reactions:
\begin{eqnarray}
	\text{C$_2$H$_4$} + \text{CH} &\to &\text{H} + \text{CH$_3$CCH} \label{eq_1a}\\
	\text{C$_3$H$_5$} + \text{H} &\to & \text{H$_2$} + \text{CH$_3$CCH} \label{eq_1b}
\end{eqnarray}

 C$_2$H$_4$ is formed in the gas phase through the CH + CH$_4$ reaction and on grains through the hydrogenation of C$_2$H$_2$, which is formed in the gas-phase. C$_3$H$_5$ is mainly formed on grains through the hydrogenation of C$_3$.
Although less efficient, the gas-phase production of CH$_3$CN is effected through the HCN + CH$_3^+$ $\rightarrow$ CH$_3$CNH$^+$ + h$\nu$ radiative association followed by the dissociative recombination of CH$_3$CNH$^+$. 

Going inward into the protostellar envelope to regions of higher density, the gas-phase formation of CH$_3$CCH is overcome by adsorption onto grain surface, explaining its abundance decrease between 2000 and 500 au. 
The small fraction of CH$_3$CN that chemically desorbs in the gas phase is preferentially consumed through several ion-neutral reactions involving H$_3^+$, HCO$^+$, He$^+$ and C$^+$.\\

\section{Discussions and conclusions}
Figures \ref{CH3CCHab} and \ref{CH3CNab} also display the observed abundances for each species (diamonds). 
For both species, the comparison between the modelled and observed abundances are based on the assumption that the H$_2$ column density has been correctly estimated for the emission region of each species.With that in mind, the model overestimates the gas-phase abundance of CH$_3$CCH by about two orders of magnitude. However, interestingly, the observed abundance seems to be located at the same radius as the modelled peak abundance. Since we use single dish observations, the observed spectra are very likely not sensitive to the innermost emission of the molecule \citep[inside 200 au, see][for a discussion on this effect]{2016MNRAS.458.1859M}, where the species abundance may be quite large as predicted by the model. Dividing the observed molecular column density by the integrated H$_2$ column density may bias the abundance towards smaller values. 
Moreover, the overestimation of CH$_3$CCH may be due to the fact that, in the model, we consider a barrier for the O + C$_3$ reaction \citep{1996ApJ...465..795W} leading to a very large C$_3$ abundance in the gas phase and then a large CH$_3$CCH abundance on grains. The overestimation of CH$_3$CCH may be an indication that the O + C$_3$ reaction 
is in fact efficient at low temperature due to tunnelling similarly to the reaction O + C$_3$H$_6$ \citep{Sabbah102}. 
For CH$_3$CN, the observed abundance could fit very well with that expected for the evaporated region. 

Using JCMT observations, \citet{2002A&A...390.1001S} have determined the abundance of these two molecules on the same source but with a different radiative transfer model and different physical properties of the source. The analysis of the observed emission was done with a "jump" model assuming a smaller constant abundance of the species in the outer part and a higher one inside 150 au where the temperature is larger than 90~K. Using this model with the observed higher frequency transitions (as compared to ours), the authors determined an inner region abundance of $7.5\times 10^{-9}$ for CH$_3$CN and $3.5\times 10^{-8}$ for CH$_3$CCH while they only had upper limits (of $8\times 10^{-11}$ for CH$_3$CN and $1.5\times 10^{-9}$ for CH$_3$CCH) for the outer part. Our chemical model predicts abundances in the outer regions that are not flat for both molecules; in particular for CH$_3$CN. Our predicted abundances for the inner regions for both molecules are very close to the ones determined by \citet{2002A&A...390.1001S}.

 Last, it is important to keep in mind that IRAS16293 is a binary system with both components inside the same observational beam for single dish observations. Variation of the chemical composition between the two binary components is currently investigated in the framework of the ALMA-PILS survey \citep{Jorgensen2016}. Although CH$_3$CN shows similar column densities towards the two components, some significant variations are observed for other species such as CH$_3$NC \citep{2018A&A...616A..90C} and C$_2$H$_3$CN \citep{2018arXiv180702909C}, possibly due to differences in their physical conditions or evolutionary stages.



\section*{Acknowledgements}

VW's research is funded by an ERC Starting Grant (3DICE, grant agreement 336474). The authors acknowledge the CNRS program "Physique et Chimie du Milieu Interstellaire" (PCMI) co-funded by the Centre National d'Etudes Spatiales (CNES). LM also acknowledges the support
from the NASA postdoctoral program. A portion of this research was
carried out at the Jet Propulsion Laboratory, California Institute
of Technology, under a contract with the National
Aeronautics and Space Administration.





\bibliographystyle{mnras}
\bibliography{bib}

\begin{thebibliography}{}
\makeatletter
\relax
\def\mn@urlcharsother{\let\do\@makeother \do\$\do\&\do\#\do\^\do\_\do\%\do\~}
\def\mn@doi{\begingroup\mn@urlcharsother \@ifnextchar [ {\mn@doi@}
  {\mn@doi@[]}}
\def\mn@doi@[#1]#2{\def\@tempa{#1}\ifx\@tempa\@empty \href
  {http://dx.doi.org/#2} {doi:#2}\else \href {http://dx.doi.org/#2} {#1}\fi
  \endgroup}
\def\mn@eprint#1#2{\mn@eprint@#1:#2::\@nil}
\def\mn@eprint@arXiv#1{\href {http://arxiv.org/abs/#1} {{\tt arXiv:#1}}}
\def\mn@eprint@dblp#1{\href {http://dblp.uni-trier.de/rec/bibtex/#1.xml}
  {dblp:#1}}
\def\mn@eprint@#1:#2:#3:#4\@nil{\def\@tempa {#1}\def\@tempb {#2}\def\@tempc
  {#3}\ifx \@tempc \@empty \let \@tempc \@tempb \let \@tempb \@tempa \fi \ifx
  \@tempb \@empty \def\@tempb {arXiv}\fi \@ifundefined
  {mn@eprint@\@tempb}{\@tempb:\@tempc}{\expandafter \expandafter \csname
  mn@eprint@\@tempb\endcsname \expandafter{\@tempc}}}

\bibitem[\protect\citeauthoryear{{Ag{\'u}ndez}, {Fonfr{\'{\i}}a}, {Cernicharo},
  {Pardo}  \& {Gu{\'e}lin}}{{Ag{\'u}ndez} et~al.}{2008}]{Agundez2008}
{Ag{\'u}ndez} M.,  {Fonfr{\'{\i}}a} J.~P.,  {Cernicharo} J.,  {Pardo} J.~R.,
  {Gu{\'e}lin} M.,  2008, \mn@doi [\aap] {10.1051/0004-6361:20078956}, \href
  {http://adsabs.harvard.edu/abs/2008A%26A...479..493A} {479, 493}

\bibitem[\protect\citeauthoryear{{Ag{\'u}ndez}, {Cernicharo},
  {Quintana-Lacaci}, {Velilla Prieto}, {Castro-Carrizo}, {Marcelino}  \&
  {Gu{\'e}lin}}{{Ag{\'u}ndez} et~al.}{2015}]{Agundez2015}
{Ag{\'u}ndez} M.,  {Cernicharo} J.,  {Quintana-Lacaci} G.,  {Velilla Prieto}
  L.,  {Castro-Carrizo} A.,  {Marcelino} N.,   {Gu{\'e}lin} M.,  2015, \mn@doi
  [\apj] {10.1088/0004-637X/814/2/143}, \href
  {http://adsabs.harvard.edu/abs/2015ApJ...814..143A} {814, 143}

\bibitem[\protect\citeauthoryear{{Aikawa}, {Wakelam}, {Garrod}  \&
  {Herbst}}{{Aikawa} et~al.}{2008}]{2008ApJ...674..984A}
{Aikawa} Y.,  {Wakelam} V.,  {Garrod} R.~T.,   {Herbst} E.,  2008, \mn@doi
  [\apj] {10.1086/524096}, \href
  {http://adsabs.harvard.edu/abs/2008ApJ...674..984A} {674, 984}

\bibitem[\protect\citeauthoryear{{Aikawa}, {Wakelam}, {Hersant}, {Garrod}  \&
  {Herbst}}{{Aikawa} et~al.}{2012}]{2012ApJ...760...40A}
{Aikawa} Y.,  {Wakelam} V.,  {Hersant} F.,  {Garrod} R.~T.,   {Herbst} E.,
  2012, \mn@doi [\apj] {10.1088/0004-637X/760/1/40}, \href
  {http://adsabs.harvard.edu/abs/2012ApJ...760...40A} {760, 40}

\bibitem[\protect\citeauthoryear{{Andr{\'e}}, {Ward-Thompson}  \&
  {Barsony}}{{Andr{\'e}} et~al.}{1993}]{Andre1993}
{Andr{\'e}} P.,  {Ward-Thompson} D.,   {Barsony} M.,  1993, \mn@doi [\apj]
  {10.1086/172425}, \href {http://adsabs.harvard.edu/abs/1993ApJ...406..122A}
  {406, 122}

\bibitem[\protect\citeauthoryear{{Askne}, {Hoglund}, {Hjalmarson}  \&
  {Irvine}}{{Askne} et~al.}{1984}]{Askne1984}
{Askne} J.,  {Hoglund} B.,  {Hjalmarson} A.,   {Irvine} W.~M.,  1984, \aap,
  \href {http://adsabs.harvard.edu/abs/1984A%26A...130..311A} {130, 311}

\bibitem[\protect\citeauthoryear{{Bertin} et~al.,}{{Bertin}
  et~al.}{2017}]{2017A&A...598A..18B}
{Bertin} M.,  et~al., 2017, \mn@doi [\aap] {10.1051/0004-6361/201629394}, \href
  {http://adsabs.harvard.edu/abs/2017A%26A...598A..18B} {598, A18}

\bibitem[\protect\citeauthoryear{{Bisschop}, {J{\o}rgensen}, {Bourke},
  {Bottinelli}  \& {van Dishoeck}}{{Bisschop} et~al.}{2008}]{Bisschop2008}
{Bisschop} S.~E.,  {J{\o}rgensen} J.~K.,  {Bourke} T.~L.,  {Bottinelli} S.,
  {van Dishoeck} E.~F.,  2008, \mn@doi [\aap] {10.1051/0004-6361:200809673},
  \href {http://adsabs.harvard.edu/abs/2008A%26A...488..959B} {488, 959}

\bibitem[\protect\citeauthoryear{{Bottinelli} et~al.,}{{Bottinelli}
  et~al.}{2004}]{Bottinelli2004}
{Bottinelli} S.,  et~al., 2004, \mn@doi [\apjl] {10.1086/426964}, \href
  {http://adsabs.harvard.edu/abs/2004ApJ...617L..69B} {617, L69}

\bibitem[\protect\citeauthoryear{{Bottinelli}, {Wakelam}, {Caux}, {Vastel},
  {Aikawa}  \& {Ceccarelli}}{{Bottinelli} et~al.}{2014}]{2014MNRAS.441.1964B}
{Bottinelli} S.,  {Wakelam} V.,  {Caux} E.,  {Vastel} C.,  {Aikawa} Y.,
  {Ceccarelli} C.,  2014, \mn@doi [\mnras] {10.1093/mnras/stu700}, \href
  {http://adsabs.harvard.edu/abs/2014MNRAS.441.1964B} {441, 1964}

\bibitem[\protect\citeauthoryear{{Calcutt} et~al.,}{{Calcutt}
  et~al.}{2018a}]{2018arXiv180702909C}
{Calcutt} H.,  et~al., 2018a, preprint, \href
  {http://adsabs.harvard.edu/abs/2018arXiv180702909C} {} (\mn@eprint {arXiv}
  {1807.02909})

\bibitem[\protect\citeauthoryear{{Calcutt} et~al.,}{{Calcutt}
  et~al.}{2018b}]{2018A&A...616A..90C}
{Calcutt} H.,  et~al., 2018b, \mn@doi [\aap] {10.1051/0004-6361/201732289},
  \href {http://adsabs.harvard.edu/abs/2018A%26A...616A..90C} {616, A90}

\bibitem[\protect\citeauthoryear{Carpenter et~al.,}{Carpenter
  et~al.}{2017}]{JSSv076i01}
Carpenter B.,  et~al., 2017, \mn@doi [Journal of Statistical Software,
  Articles] {10.18637/jss.v076.i01}, 76, 1

\bibitem[\protect\citeauthoryear{{Caux} et~al.,}{{Caux}
  et~al.}{2011}]{2011A&A...532A..23C}
{Caux} E.,  et~al., 2011, \mn@doi [\aap] {10.1051/0004-6361/201015399}, \href
  {http://adsabs.harvard.edu/abs/2011A%26A...532A..23C} {532, A23}

\bibitem[\protect\citeauthoryear{{Cazaux}, {Tielens}, {Ceccarelli}, {Castets},
  {Wakelam}, {Caux}, {Parise}  \& {Teyssier}}{{Cazaux}
  et~al.}{2003}]{Cazaux2003}
{Cazaux} S.,  {Tielens} A.~G.~G.~M.,  {Ceccarelli} C.,  {Castets} A.,
  {Wakelam} V.,  {Caux} E.,  {Parise} B.,   {Teyssier} D.,  2003, \mn@doi
  [\apjl] {10.1086/378038}, \href
  {http://adsabs.harvard.edu/abs/2003ApJ...593L..51C} {593, L51}

\bibitem[\protect\citeauthoryear{{Cazzoli} \& {Puzzarini}}{{Cazzoli} \&
  {Puzzarini}}{2008}]{Cazzoli.2008}
{Cazzoli} G.,  {Puzzarini} C.,  2008, \mn@doi [\aap]
  {10.1051/0004-6361:200809938}, \href
  {http://cdsads.u-strasbg.fr/abs/2008A%26A...487.1197C} {487, 1197}

\bibitem[\protect\citeauthoryear{{Crimier}, {Ceccarelli}, {Maret},
  {Bottinelli}, {Caux}, {Kahane}, {Lis}  \& {Olofsson}}{{Crimier}
  et~al.}{2010}]{2010A&A...519A..65C}
{Crimier} N.,  {Ceccarelli} C.,  {Maret} S.,  {Bottinelli} S.,  {Caux} E.,
  {Kahane} C.,  {Lis} D.~C.,   {Olofsson} J.,  2010, \mn@doi [\aap]
  {10.1051/0004-6361/200913112}, \href
  {http://adsabs.harvard.edu/abs/2010A%26A...519A..65C} {519, A65}

\bibitem[\protect\citeauthoryear{{Dickens}, {Irvine}, {Snell}, {Bergin},
  {Schloerb}, {Pratap}  \& {Miralles}}{{Dickens}
  et~al.}{2000}]{2000ApJ...542..870D}
{Dickens} J.~E.,  {Irvine} W.~M.,  {Snell} R.~L.,  {Bergin} E.~A.,  {Schloerb}
  F.~P.,  {Pratap} P.,   {Miralles} M.~P.,  2000, \mn@doi [\apj]
  {10.1086/317040}, \href {http://adsabs.harvard.edu/abs/2000ApJ...542..870D}
  {542, 870}

\bibitem[\protect\citeauthoryear{{Dzib} et~al.,}{{Dzib}
  et~al.}{2018}]{2018arXiv180203234D}
{Dzib} S.~A.,  et~al., 2018, preprint, \href
  {http://adsabs.harvard.edu/abs/2018arXiv180203234D} {} (\mn@eprint {arXiv}
  {1802.03234})

\bibitem[\protect\citeauthoryear{{Fayolle}, {{\"O}berg}, {Garrod}, {van
  Dishoeck}  \& {Bisschop}}{{Fayolle} et~al.}{2015}]{Fayolle2015}
{Fayolle} E.~C.,  {{\"O}berg} K.~I.,  {Garrod} R.~T.,  {van Dishoeck} E.~F.,
  {Bisschop} S.~E.,  2015, \mn@doi [\aap] {10.1051/0004-6361/201323114}, \href
  {http://adsabs.harvard.edu/abs/2015A%26A...576A..45F} {576, A45}

\bibitem[\protect\citeauthoryear{{Garrod}, {Wakelam}  \& {Herbst}}{{Garrod}
  et~al.}{2007}]{2007A&A...467.1103G}
{Garrod} R.~T.,  {Wakelam} V.,   {Herbst} E.,  2007, \mn@doi [\aap]
  {10.1051/0004-6361:20066704}, \href
  {http://adsabs.harvard.edu/abs/2007A%26A...467.1103G} {467, 1103}

\bibitem[\protect\citeauthoryear{Gelman \& Rubin}{Gelman \&
  Rubin}{1992}]{Gelman.1992}
Gelman A.,  Rubin D.,  1992, Statistical Science, 7, 457

\bibitem[\protect\citeauthoryear{{Graedel}, {Langer}  \& {Frerking}}{{Graedel}
  et~al.}{1982}]{1982ApJS...48..321G}
{Graedel} T.~E.,  {Langer} W.~D.,   {Frerking} M.~A.,  1982, \mn@doi [\apjs]
  {10.1086/190780}, \href {http://adsabs.harvard.edu/abs/1982ApJS...48..321G}
  {48, 321}

\bibitem[\protect\citeauthoryear{{Gratier}, {Pety}, {Guzm{\'a}n}, {Gerin},
  {Goicoechea}, {Roueff}  \& {Faure}}{{Gratier} et~al.}{2013}]{Gratier2013}
{Gratier} P.,  {Pety} J.,  {Guzm{\'a}n} V.,  {Gerin} M.,  {Goicoechea} J.~R.,
  {Roueff} E.,   {Faure} A.,  2013, \mn@doi [\aap]
  {10.1051/0004-6361/201321031}, \href
  {http://adsabs.harvard.edu/abs/2013A%26A...557A.101G} {557, A101}

\bibitem[\protect\citeauthoryear{{Gratier}, {Majumdar}, {Ohishi}, {Roueff},
  {Loison}, {Hickson}  \& {Wakelam}}{{Gratier}
  et~al.}{2016}]{2016ApJS..225...25G}
{Gratier} P.,  {Majumdar} L.,  {Ohishi} M.,  {Roueff} E.,  {Loison} J.~C.,
  {Hickson} K.~M.,   {Wakelam} V.,  2016, \mn@doi [\apjs]
  {10.3847/0067-0049/225/2/25}, \href
  {http://adsabs.harvard.edu/abs/2016ApJS..225...25G} {225, 25}

\bibitem[\protect\citeauthoryear{{Green}}{{Green}}{1986}]{1986ApJ...309..331G}
{Green} S.,  1986, \mn@doi [\apj] {10.1086/164605}, \href
  {https://ui.adsabs.harvard.edu/#abs/1986ApJ...309..331G} {309, 331}

\bibitem[\protect\citeauthoryear{{Guzm{\'a}n}, {Pety}, {Gratier}, {Goicoechea},
  {Gerin}, {Roueff}, {Le Petit}  \& {Le Bourlot}}{{Guzm{\'a}n}
  et~al.}{2014}]{Guzman2014}
{Guzm{\'a}n} V.~V.,  {Pety} J.,  {Gratier} P.,  {Goicoechea} J.~R.,  {Gerin}
  M.,  {Roueff} E.,  {Le Petit} F.,   {Le Bourlot} J.,  2014, \mn@doi [Faraday
  Discussions] {10.1039/c3fd00114h}, \href
  {http://adsabs.harvard.edu/abs/2014FaDi..168..103G} {168, 103}

\bibitem[\protect\citeauthoryear{{Hasegawa} \& {Herbst}}{{Hasegawa} \&
  {Herbst}}{1993}]{1993MNRAS.261...83H}
{Hasegawa} T.~I.,  {Herbst} E.,  1993, \mn@doi [\mnras]
  {10.1093/mnras/261.1.83}, \href
  {http://adsabs.harvard.edu/abs/1993MNRAS.261...83H} {261, 83}

\bibitem[\protect\citeauthoryear{{Herbst} \& {van Dishoeck}}{{Herbst} \& {van
  Dishoeck}}{2009}]{Herbst2009}
{Herbst} E.,  {van Dishoeck} E.~F.,  2009, \mn@doi [\araa]
  {10.1146/annurev-astro-082708-101654}, \href
  {http://adsabs.harvard.edu/abs/2009ARA%26A..47..427H} {47, 427}

\bibitem[\protect\citeauthoryear{{Hickson}, {Wakelam}  \& {Loison}}{{Hickson}
  et~al.}{2016}]{2016MolAs...3....1H}
{Hickson} K.~M.,  {Wakelam} V.,   {Loison} J.-C.,  2016, \mn@doi [Molecular
  Astrophysics] {10.1016/j.molap.2016.03.001}, \href
  {http://adsabs.harvard.edu/abs/2016MolAs...3....1H} {3, 1}

\bibitem[\protect\citeauthoryear{{Hincelin}, {Wakelam}, {Hersant},
  {Guilloteau}, {Loison}, {Honvault}  \& {Troe}}{{Hincelin}
  et~al.}{2011}]{2011A&A...530A..61H}
{Hincelin} U.,  {Wakelam} V.,  {Hersant} F.,  {Guilloteau} S.,  {Loison} J.~C.,
   {Honvault} P.,   {Troe} J.,  2011, \mn@doi [\aap]
  {10.1051/0004-6361/201016328}, \href
  {http://adsabs.harvard.edu/abs/2011A%26A...530A..61H} {530, A61}

\bibitem[\protect\citeauthoryear{{Jacobsen} et~al.,}{{Jacobsen}
  et~al.}{2018}]{2018A&A...612A..72J}
{Jacobsen} S.~K.,  et~al., 2018, \mn@doi [\aap] {10.1051/0004-6361/201731668},
  \href {http://adsabs.harvard.edu/abs/2018A%26A...612A..72J} {612, A72}

\bibitem[\protect\citeauthoryear{{Jenkins}}{{Jenkins}}{2009}]{2009ApJ...700.1299J}
{Jenkins} E.~B.,  2009, \mn@doi [\apj] {10.1088/0004-637X/700/2/1299}, \href
  {http://adsabs.harvard.edu/abs/2009ApJ...700.1299J} {700, 1299}

\bibitem[\protect\citeauthoryear{{J{\o}rgensen}, {Bourke}, {Nguyen Luong}  \&
  {Takakuwa}}{{J{\o}rgensen} et~al.}{2011}]{Jorgensen2011}
{J{\o}rgensen} J.~K.,  {Bourke} T.~L.,  {Nguyen Luong} Q.,   {Takakuwa} S.,
  2011, \mn@doi [\aap] {10.1051/0004-6361/201117139}, \href
  {http://adsabs.harvard.edu/abs/2011A%26A...534A.100J} {534, A100}

\bibitem[\protect\citeauthoryear{{J{\o}rgensen}, {Favre}, {Bisschop}, {Bourke},
  {van Dishoeck}  \& {Schmalzl}}{{J{\o}rgensen} et~al.}{2012}]{Jorgensen2012}
{J{\o}rgensen} J.~K.,  {Favre} C.,  {Bisschop} S.~E.,  {Bourke} T.~L.,  {van
  Dishoeck} E.~F.,   {Schmalzl} M.,  2012, \mn@doi [\apjl]
  {10.1088/2041-8205/757/1/L4}, \href
  {http://adsabs.harvard.edu/abs/2012ApJ...757L...4J} {757, L4}

\bibitem[\protect\citeauthoryear{{J{\o}rgensen} et~al.,}{{J{\o}rgensen}
  et~al.}{2016}]{Jorgensen2016}
{J{\o}rgensen} J.~K.,  et~al., 2016, \mn@doi [\aap]
  {10.1051/0004-6361/201628648}, \href
  {http://adsabs.harvard.edu/abs/2016A%26A...595A.117J} {595, A117}

\bibitem[\protect\citeauthoryear{{Kahane}, {Ceccarelli}, {Faure}  \&
  {Caux}}{{Kahane} et~al.}{2013}]{Kahane2013}
{Kahane} C.,  {Ceccarelli} C.,  {Faure} A.,   {Caux} E.,  2013, \mn@doi [\apjl]
  {10.1088/2041-8205/763/2/L38}, \href
  {http://adsabs.harvard.edu/abs/2013ApJ...763L..38K} {763, L38}

\bibitem[\protect\citeauthoryear{{Kalenskii}, {Promislov}, {Alakoz}, {Winnberg}
   \& {Johansson}}{{Kalenskii} et~al.}{2000}]{Kalenskii2000}
{Kalenskii} S.~V.,  {Promislov} V.~G.,  {Alakoz} A.,  {Winnberg} A.~V.,
  {Johansson} L.~E.~B.,  2000, \aap, \href
  {http://adsabs.harvard.edu/abs/2000A%26A...354.1036K} {354, 1036}

\bibitem[\protect\citeauthoryear{{Ligterink} et~al.,}{{Ligterink}
  et~al.}{2017}]{Ligterink2017}
{Ligterink} N.~F.~W.,  et~al., 2017, \mn@doi [\mnras] {10.1093/mnras/stx890},
  \href {http://adsabs.harvard.edu/abs/2017MNRAS.469.2219L} {469, 2219}

\bibitem[\protect\citeauthoryear{{Lykke} et~al.,}{{Lykke}
  et~al.}{2017}]{Lykke2017}
{Lykke} J.~M.,  et~al., 2017, \mn@doi [\aap] {10.1051/0004-6361/201629180},
  \href {http://adsabs.harvard.edu/abs/2017A%26A...597A..53L} {597, A53}

\bibitem[\protect\citeauthoryear{{Majumdar}, {Gratier}, {Vidal}, {Wakelam},
  {Loison}, {Hickson}  \& {Caux}}{{Majumdar}
  et~al.}{2016}]{2016MNRAS.458.1859M}
{Majumdar} L.,  {Gratier} P.,  {Vidal} T.,  {Wakelam} V.,  {Loison} J.-C.,
  {Hickson} K.~M.,   {Caux} E.,  2016, \mn@doi [\mnras] {10.1093/mnras/stw457},
  \href {http://adsabs.harvard.edu/abs/2016MNRAS.458.1859M} {458, 1859}

\bibitem[\protect\citeauthoryear{{Majumdar}, {Gratier}, {Andron}, {Wakelam}  \&
  {Caux}}{{Majumdar} et~al.}{2017}]{2017MNRAS.467.3525M}
{Majumdar} L.,  {Gratier} P.,  {Andron} I.,  {Wakelam} V.,   {Caux} E.,  2017,
  \mn@doi [\mnras] {10.1093/mnras/stx259}, \href
  {http://adsabs.harvard.edu/abs/2017MNRAS.467.3525M} {467, 3525}

\bibitem[\protect\citeauthoryear{{Majumdar}, {Gratier}, {Wakelam}, {Caux},
  {Willacy}  \& {Ressler}}{{Majumdar} et~al.}{2018}]{2018arXiv180305442M}
{Majumdar} L.,  {Gratier} P.,  {Wakelam} V.,  {Caux} E.,  {Willacy} K.,
  {Ressler} M.~E.,  2018, \mn@doi [\mnras] {10.1093/mnras/sty703}, \href
  {https://ui.adsabs.harvard.edu/#abs/2018MNRAS.477..525M} {477, 525}

\bibitem[\protect\citeauthoryear{{Masunaga} \& {Inutsuka}}{{Masunaga} \&
  {Inutsuka}}{2000}]{2000ApJ...531..350M}
{Masunaga} H.,  {Inutsuka} S.-i.,  2000, \mn@doi [\apj] {10.1086/308439}, \href
  {http://adsabs.harvard.edu/abs/2000ApJ...531..350M} {531, 350}

\bibitem[\protect\citeauthoryear{{Mauersberger}, {Henkel}, {Walmsley}, {Sage}
  \& {Wiklind}}{{Mauersberger} et~al.}{1991}]{Mauersberger1991}
{Mauersberger} R.,  {Henkel} C.,  {Walmsley} C.~M.,  {Sage} L.~J.,   {Wiklind}
  T.,  1991, \aap, \href {http://adsabs.harvard.edu/abs/1991A%26A...247..307M}
  {247, 307}

\bibitem[\protect\citeauthoryear{{M{\"u}ller}, {Schl{\"o}der}, {Stutzki}  \&
  {Winnewisser}}{{M{\"u}ller} et~al.}{2005}]{2005JMoSt.742..215M}
{M{\"u}ller} H.~S.~P.,  {Schl{\"o}der} F.,  {Stutzki} J.,   {Winnewisser} G.,
  2005, \mn@doi [Journal of Molecular Structure]
  {10.1016/j.molstruc.2005.01.027}, \href
  {http://adsabs.harvard.edu/abs/2005JMoSt.742..215M} {742, 215}

\bibitem[\protect\citeauthoryear{{M{\"u}ller} et~al.,}{{M{\"u}ller}
  et~al.}{2015}]{Muller.2015}
{M{\"u}ller} H.~S.~P.,  et~al., 2015, \mn@doi [Journal of Molecular
  Spectroscopy] {10.1016/j.jms.2015.02.009}, \href
  {http://cdsads.u-strasbg.fr/abs/2015JMoSp.312...22M} {312, 22}

\bibitem[\protect\citeauthoryear{{Neufeld}, {Wolfire}  \& {Schilke}}{{Neufeld}
  et~al.}{2005}]{2005ApJ...628..260N}
{Neufeld} D.~A.,  {Wolfire} M.~G.,   {Schilke} P.,  2005, \mn@doi [\apj]
  {10.1086/430663}, \href {http://adsabs.harvard.edu/abs/2005ApJ...628..260N}
  {628, 260}

\bibitem[\protect\citeauthoryear{{{\"O}berg}, {Guzm{\'a}n}, {Furuya}, {Qi},
  {Aikawa}, {Andrews}, {Loomis}  \& {Wilner}}{{{\"O}berg}
  et~al.}{2015}]{Oberg2015}
{{\"O}berg} K.~I.,  {Guzm{\'a}n} V.~V.,  {Furuya} K.,  {Qi} C.,  {Aikawa} Y.,
  {Andrews} S.~M.,  {Loomis} R.,   {Wilner} D.~J.,  2015, \mn@doi [\nat]
  {10.1038/nature14276}, \href
  {http://adsabs.harvard.edu/abs/2015Natur.520..198O} {520, 198}

\bibitem[\protect\citeauthoryear{{Ortiz-Le{\'o}n} et~al.,}{{Ortiz-Le{\'o}n}
  et~al.}{2017}]{Ortiz2017}
{Ortiz-Le{\'o}n} G.~N.,  et~al., 2017, \mn@doi [\apj]
  {10.3847/1538-4357/834/2/141}, \href
  {http://adsabs.harvard.edu/abs/2017ApJ...834..141O} {834, 141}

\bibitem[\protect\citeauthoryear{{Podio} et~al.,}{{Podio}
  et~al.}{2015}]{2015A&A...581A..85P}
{Podio} L.,  et~al., 2015, \mn@doi [\aap] {10.1051/0004-6361/201525778}, \href
  {http://adsabs.harvard.edu/abs/2015A%26A...581A..85P} {581, A85}

\bibitem[\protect\citeauthoryear{{Ruaud}, {Wakelam}  \& {Hersant}}{{Ruaud}
  et~al.}{2016}]{2016MNRAS.459.3756R}
{Ruaud} M.,  {Wakelam} V.,   {Hersant} F.,  2016, \mn@doi [\mnras]
  {10.1093/mnras/stw887}, \href
  {http://adsabs.harvard.edu/abs/2016MNRAS.459.3756R} {459, 3756}

\bibitem[\protect\citeauthoryear{Sabbah, Biennier, Sims, Georgievskii,
  Klippenstein  \& Smith}{Sabbah et~al.}{2007}]{Sabbah102}
Sabbah H.,  Biennier L.,  Sims I.~R.,  Georgievskii Y.,  Klippenstein S.~J.,
  Smith I. W.~M.,  2007, \mn@doi [Science] {10.1126/science.1142373}, 317, 102

\bibitem[\protect\citeauthoryear{{Sch{\"o}ier}, {J{\o}rgensen}, {van Dishoeck}
  \& {Blake}}{{Sch{\"o}ier} et~al.}{2002}]{2002A&A...390.1001S}
{Sch{\"o}ier} F.~L.,  {J{\o}rgensen} J.~K.,  {van Dishoeck} E.~F.,   {Blake}
  G.~A.,  2002, \mn@doi [\aap] {10.1051/0004-6361:20020756}, \href
  {http://adsabs.harvard.edu/abs/2002A%26A...390.1001S} {390, 1001}

\bibitem[\protect\citeauthoryear{{Vastel}, {Ceccarelli}, {Lefloch}  \&
  {Bachiller}}{{Vastel} et~al.}{2014}]{Vastel2014}
{Vastel} C.,  {Ceccarelli} C.,  {Lefloch} B.,   {Bachiller} R.,  2014, \mn@doi
  [\apjl] {10.1088/2041-8205/795/1/L2}, \href
  {http://adsabs.harvard.edu/abs/2014ApJ...795L...2V} {795, L2}

\bibitem[\protect\citeauthoryear{{Vidal} \& {Wakelam}}{{Vidal} \&
  {Wakelam}}{2018}]{2018MNRAS.474.5575V}
{Vidal} T.~H.~G.,  {Wakelam} V.,  2018, \mn@doi [\mnras]
  {10.1093/mnras/stx3113}, \href
  {http://adsabs.harvard.edu/abs/2018MNRAS.474.5575V} {474, 5575}

\bibitem[\protect\citeauthoryear{{Vidal}, {Loison}, {Jaziri}, {Ruaud},
  {Gratier}  \& {Wakelam}}{{Vidal} et~al.}{2017}]{2017MNRAS.469..435V}
{Vidal} T.~H.~G.,  {Loison} J.-C.,  {Jaziri} A.~Y.,  {Ruaud} M.,  {Gratier} P.,
    {Wakelam} V.,  2017, \mn@doi [\mnras] {10.1093/mnras/stx828}, \href
  {http://adsabs.harvard.edu/abs/2017MNRAS.469..435V} {469, 435}

\bibitem[\protect\citeauthoryear{{Viti}, {Caselli}, {Hartquist}  \&
  {Williams}}{{Viti} et~al.}{2001}]{2001A&A...370.1017V}
{Viti} S.,  {Caselli} P.,  {Hartquist} T.~W.,   {Williams} D.~A.,  2001,
  \mn@doi [\aap] {10.1051/0004-6361:20010300}, \href
  {http://adsabs.harvard.edu/abs/2001A%26A...370.1017V} {370, 1017}

\bibitem[\protect\citeauthoryear{{Wakelam} \& {Herbst}}{{Wakelam} \&
  {Herbst}}{2008}]{2008ApJ...680..371W}
{Wakelam} V.,  {Herbst} E.,  2008, \mn@doi [\apj] {10.1086/587734}, \href
  {http://adsabs.harvard.edu/abs/2008ApJ...680..371W} {680, 371}

\bibitem[\protect\citeauthoryear{{Wakelam}, {Vastel}, {Aikawa}, {Coutens},
  {Bottinelli}  \& {Caux}}{{Wakelam} et~al.}{2014}]{2014MNRAS.445.2854W}
{Wakelam} V.,  {Vastel} C.,  {Aikawa} Y.,  {Coutens} A.,  {Bottinelli} S.,
  {Caux} E.,  2014, \mn@doi [\mnras] {10.1093/mnras/stu1920}, \href
  {http://adsabs.harvard.edu/abs/2014MNRAS.445.2854W} {445, 2854}

\bibitem[\protect\citeauthoryear{{Wakelam} et~al.,}{{Wakelam}
  et~al.}{2015}]{2015ApJS..217...20W}
{Wakelam} V.,  et~al., 2015, \mn@doi [\apjs] {10.1088/0067-0049/217/2/20},
  \href {http://adsabs.harvard.edu/abs/2015ApJS..217...20W} {217, 20}

\bibitem[\protect\citeauthoryear{{Wakelam}, {Loison}, {Mereau}  \&
  {Ruaud}}{{Wakelam} et~al.}{2017}]{2017MolAs...6...22W}
{Wakelam} V.,  {Loison} J.-C.,  {Mereau} R.,   {Ruaud} M.,  2017, \mn@doi
  [Molecular Astrophysics] {10.1016/j.molap.2017.01.002}, \href
  {http://adsabs.harvard.edu/abs/2017MolAs...6...22W} {6, 22}

\bibitem[\protect\citeauthoryear{{Woon} \& {Herbst}}{{Woon} \&
  {Herbst}}{1996}]{1996ApJ...465..795W}
{Woon} D.~E.,  {Herbst} E.,  1996, \mn@doi [\apj] {10.1086/177463}, \href
  {http://adsabs.harvard.edu/abs/1996ApJ...465..795W} {465, 795}

\bibitem[\protect\citeauthoryear{{van Dishoeck}, {Blake}, {Jansen}  \&
  {Groesbeck}}{{van Dishoeck} et~al.}{1995}]{vanDishoeck1995}
{van Dishoeck} E.~F.,  {Blake} G.~A.,  {Jansen} D.~J.,   {Groesbeck} T.~D.,
  1995, \mn@doi [\apj] {10.1086/175915}, \href
  {http://adsabs.harvard.edu/abs/1995ApJ...447..760V} {447, 760}

\makeatother
\end{thebibliography}







\bsp	
\label{lastpage}
\end{document}